\documentclass[aps,prl,twocolumn,nofootinbib,superscriptaddress,amssymb]{revtex4}
\usepackage{color}
\usepackage{epsfig}
\usepackage{graphicx}
\usepackage{epstopdf}
\usepackage{mathtools}
\usepackage{bbold}

\def\be{\begin{equation}}
\def\ee{\end{equation}}

\begin{document}

\title{Behind the Horizon in AdS/CFT}

\author{Erik Verlinde}

\affiliation{Institute for Theoretical Physics, University of Amsterdam, Amsterdam, The Netherlands}

\def\spc{\hspace{.5pt}}

\author{Herman Verlinde}
\affiliation{Department of Physics, Princeton University, Princeton, NJ 08544, USA}

\date{\today}

\begin{abstract}
We extend the recent proposal of \cite{PR} of a CFT construction of operators inside the black hole interior to arbitrary non-maximally mixed states. Our construction builds on the general prescription given  in the earlier work  \cite{vv}.  We indicate how the CFT state dependence of the interior modes can be removed by introducing an external system, such as an observer, that is entangled with the CFT.

\end{abstract}

\def\calO{{b}}
\def\be{\begin{equation}}
\def\ee{\end{equation}}



\maketitle
\def\mathbi#1{\textbf{\em #1}} 
\def\som{{ \textit{\textbf s}}} 
\def\tom{{ \textit{\textbf t}}} 
\def\nom{n} 
\def\mom{m} 
\def\la{\langle}
\def\bea{\begin{eqnarray}}
\def\eea{\end{eqnarray}}
\def\is{\! & \! = \! & \!}
\def\ra{\rangle}
\def\half{{\textstyle{\frac 12}}}
\def\cL{{\cal L}}
\def\halfi{{\textstyle{\frac i 2}}}
\def\ba{\begin{eqnarray}}
\def\ea{\end{eqnarray}}





\def\ibar{\spc f\spc} 
\def\kbar{{\spc {\bar{k}\spc }}}

\newcommand{\rep}[1]{\mathbf{#1}}
\newcommand{\Tr}{\, {\rm Tr}}
\def\uU{\bf U}
\def\be{\bea}
\def\ee{\eea}
\def\delbar{\overline{\partial}}
\newcommand{\smpc}{\hspace{.5pt}}
\def\ra{\bigr\rangle}
\def\la{\bigl\langle}
\def\ccdot{\!\spc\cdot\!\spc}
\def\nspc{\!\spc\smpc}
\def\tr{{\rm tr}}
\def\bh{{\mbox{\fontsize{7pt}{.7pt}{$BH$}}}}

\def\Aa{{\mbox{\scriptsize \smpc \sc a}}}
\def\bfC{\mbox{{\textbf C}}}
\def\bC{\alpha} 
\def\nonu{\nonumber}
\def\sC{{\mbox{\scriptsize {\smpc \sc sc}}}}
\addtolength{\baselineskip}{.3mm}
\addtolength{\parskip}{.3mm}
\renewcommand\Large{\fontsize{15.5}{16}\selectfont}
\def\ra{\bigr\rangle}
\def\la{\bigl\langle}
\def\li{\bigl|\spc}
\def\ri{\bigr |\spc}

\def\hf{\textstyle \frac 1 2}

\def\cE{{\mbox{\tiny \nspc $\callE$}}}
\def\Ee{{\mbox{\scriptsize \smpc \sc e}}}
\def\Bb{{\raisebox{-.2pt}{\scriptsize \smpc \sc b}}}
\def\Zz{{\raisebox{-.2pt}{\scriptsize \smpc \sc z}}}

\def\Bbt{{\raisebox{-.2pt}{\scriptsize \smpc $\tilde{\mbox{\sc b}}$}}}
\def\Hh{{\mbox{\scriptsize \smpc \sc h}}}
\def\Aa{{\mbox{\scriptsize \smpc \sc a}}}

\def\AB{{\mbox{\scriptsize \smpc \sc ab}}}

\def\BH{{\mbox{\scriptsize \smpc \sc bh}}}

\def\AE{{\mbox{\scriptsize \smpc \sc ae}}}

\def\callE{{\mbox{E}}}
\def\callBH{{\mbox{BH}}}

\def\callH{{\mbox{H}}}
\def\callB{{\mbox{B}}}
\setcounter{tocdepth}{2}
\addtolength{\baselineskip}{.0mm}
\addtolength{\parskip}{.0mm}
\addtolength{\abovedisplayskip}{.2mm}
\addtolength{\belowdisplayskip}{.2mm}

\subsection{Introduction} 
\vspace{-3mm}

The question whether AdS/CFT admits the construction of operators that describe the interior of a black hole  has been a focus of attention over the past year \cite{amps,PR,MP}. In a recent paper \cite{PR},  Papadodimas and Raju proposed an elegant prescription, that deserves further scrutiny.

The PR proposal employs the CFT realization of field operators in the exterior geometry \cite{BDHM}
\bea
\label{field}
\phi_{{}_{\! \rm CFT}}(t,\Omega, z) = \sum_{\ell, \; \omega>0}\, \bigl(\calO_{\ell,\omega} f_{\ell, \omega}(z,\Omega,t) + {\rm h.c.} \bigr)
\eea
where $f_{\ell, \omega}(z,\Omega,t)$ are exterior mode functions and $\calO_{\ell,\omega}$ are
single trace operators in the CFT, that in the large N limit satisfy the free field commutation relations
\bea
\label{cancom}
\bigl[\spc \calO_\omega, \spc  \calO^\dag_\xi\spc \bigr]    \is \delta_{\omega \xi}.
\eea
Here, and in the following, we suppress all quantum numbers of the mode operators $\calO$ except for the frequency~$\omega$. 

Let $\li \Psi\ra$ denote some typical equilibrium CFT state
with total energy $E$. Expectation values of external operators $\calO_\omega$ with respect to $\li \Psi \ra$ look like thermal expectation values with inverse temperature~$\beta$. 
AdS/CFT duality identifies $\li \Psi \ra$  with the state of an AdS black hole with mass $M=E$. Following \cite{PR}, we may then define the internal operators $\widetilde\calO_\omega$ and $\widetilde\calO^\dag_\omega$ via the mirror property 
\bea
\label{mirrorone}
\widetilde\calO_\omega A_\alpha \li \Psi \ra 
\is e^{-\frac{\beta}{2}  \omega}  A_\alpha\,\calO^\dag_\omega   \li \Psi\ra \\[3mm]
\widetilde\calO^\dag_\omega A_\alpha \li \Psi \ra 
\is e^{\frac{\beta}{2}  \omega}  A_\alpha\,\calO_\omega   \li \Psi\ra 
\label{mirrortwo}
\eea
where $A_\alpha$ denotes an arbitrary polynomical of the external $\calO_\omega$ operators.
PR then proceed to give a state dependent construction of the mirror operators via \cite{PR}:
\bea
\label{prprop}
\widetilde\calO_\omega^{PR} \! = g^{mn} | u_m \rangle \langle v_n|,\ \qquad\  g_{mn}\! = \nspc \langle v_m |  v_n\rangle\nonumber\\[-1.5mm]\\[-1.5mm]
| v_n \rangle = A_n | \Psi\rangle,\  \qquad \  | u_n\rangle= e^{-\frac{\beta} 2 \omega} A_n \calO^\dag_\omega | \Psi \rangle\nonumber
\eea
where $A_n$ denotes a complete basis of operators made out of the $\calO_\omega$ oscillators.
The PR construction works for pure states, but introduces a non-linear state dependence.\footnote{Other somewhat unsatisfactory aspects of the PR mirror map is that it postulates the Unruh form of the CFT state as an exact rather than a statistical property and does not manifestly commute with hermitian conjugation. The requirement that the mirror of the adjoint 
equals the adjoint of the mirror is upheld, but only within exponential accuracy, thanks to the KMS property of the equilibrium state~$|\Psi\rangle$. }

This non-linear state dependence is  rather worrisome.  Suppose we add an 
external system X that interacts with the CFT, in such a way that asymptotic AdS modes can leak out. 
This is a model of black hole evaporation. Over time, the CFT then  evolves into a 
mixed state, that is entangled with the external system
\be
\label{entangled}
\li \Psi\ra = \sum_i\;  \li \Psi_i \ra_{\rm CFT} \; \li \psi_i \ra_{\rm X}
\ee
The standard rules of quantum mechanics require that CFT observables can act on typical entangled states of this form. This means that they have to act linearly on each state  $| \Psi_i \rangle_{\rm CFT}$
that appears in the sum (\ref{entangled}). The definition (\ref{prprop}) therefore does not work for mixed CFT states like (\ref{entangled}), and needs to be generalized.

In this note, we give a definition of interior operators for mixed CFT states with an arbitrary density matrix 
\bea
\label{mixed}
\rho = \sum_{ij} \rho_{ij} \li \Psi_i \ra \la \Psi_j \ri.
\eea Our construction is a direct application to AdS/CFT of the prescription given in \cite{vv}, and works 
for any non-maximally mixed state, with von Neumann entropy $S_{\rm vN} = - \tr(\rho\log \rho)$ sufficiently less than the maximal entropy $S_{\rm BH}$.
In our set up, we are also forced to introduce some state dependence, but of a relatively mild form.  
The discussion below will be succinct. More detailed explanations can be found in \cite{vv}.

\subsection{Equilibrium State} 
\vspace{-3mm}

Pick a collection of modes $\calO_\omega$ that span the 
free field Hilbert space of the outside region, extended over some large but finite time interval and cut-off at the stretched horizon.
Consider the class of special CFT states $\li \Psi_0\ra$ of energy $E$ that (to very high accuracy) are annihilated by all 
external lowering operators.
\bea
\label{vacco}
\calO_{\omega}\li  \Psi_0 \ra \is 0, \qquad \forall \, \omega.
\eea
This  condition selects states that do not contain any single trace excitations in the chosen time interval.
As long as $E\gg \sum_\omega \omega$, there are many such states $\li \Psi_0\ra$.
They  are all non-equilibrium states: from the bulk perspective, they
describe black holes with a singular horizon, with boundary conditions prescribed by the Boulware vacuum.
We can think of the $\li \Psi_0\ra$ as eigen states of  the deformed CFT Hamiltonian $H(\lambda) = H_{\rm CFT} + \lambda \sum_\omega \omega N_\omega$, with $N_\omega$ the number operator of $\calO_\omega$, with the property that the energy $E$  remains finite in the large $\lambda$ limit. 

By repeatedly acting with single trace raising operators $\calO^\dag_\omega$ on the vacuum state $|\Psi_0\rangle$, we can now build a free field Fock space with orthonormal basis states 
\bea
\li n \ra = A_n \li \Psi_0\ra, \ \ & & \ \
 \la n\ri m \ra 
 = \delta_{nm}\,. 
\eea
To leading order in $1/N$, the CFT Hilbert space decomposes into a direct sum ${\cal H}_{\rm CFT} = \oplus_i {\cal H}_i$ of distinct copies  of the $b_\omega$ Fock space, with each sector ${\cal H}_i$ built on a different vacuum state $| \, i \, \rangle$ with $b_\omega|\, i\,\rangle = 0$. As emphasized in \cite{PR}, it is reasonable to assume that at finite $N$,  the label $n$ runs over a large but finite set of basis states.

Any general outside operator ${\cal O}$, given by some polynomial in the $\calO^\dag$ and $\calO$ oscillators, can be characterized by its action on the basis operators $A_n$ via
\bea
\label{matact}
\hat{\cal O} \, A_n = \sum_{m} \, {\cal O}_{nm} \, A_m
\eea
where the coefficients ${\cal O}_{nm} = \langle m | \hat{\cal O} | n\rangle$ are real c-number matrix elements.
We will use this notation momentarily.

Now consider the following `quantum quench' scenario \cite{CC}: imagine that for $t<0$, the  
Hamiltonian is given by deformed CFT Hamiltonian $H(\lambda)$ with $\lambda \gg1$. At $t=0$, we suddenly turn off the deformation: we set $\lambda = 0$ and let time evolve with $H = H_{\rm CFT}$. Each non-equilibrium state $\li \Psi_0\ra$ then relaxes into a thermal equilibrium state
\bea
\label{evol}
\li \Psi \ra \is e^{ - i \tau H} \li \Psi_0 \ra 
\eea
where $\tau$ denotes the relaxation time. $\tau$ is expected to be of order of a few times the scrambling time.

The above discussion mirrors the set up in \cite{vv}.  Next, following \cite{vv},  we use the fact that $|\Psi \rangle$ can always be decomposed as follows 
\bea
\label{kdeco}
\li \Psi \ra = \sum_{n>0} A_n  {\bf C}_n \li \Psi_0 \ra
\eea
where, to the required accuracy, ${\bf C}_n$ satisfy the properties
\bea
\label{kdef}
\bigl[{\bf C}_n , A_m \bigr] = 0, \qquad \sum_n {\bf C}_n^\dag {\bf C}_n = \mathbb{1}\, .
\eea
Equation (\ref{kdeco}) expresses $| \Psi\rangle$ as a sum of terms,  given by a factor in the $\calO^\dag$ Fock space, times a factor in the sector of all $\calO$ vacua.  The ${\bf C}_n$ act within this vacuum sector, and thus  commute (to leading order in $1/N$) with all operators that act within the $\calO^\dag$ Fock space. 
Physically, we can interpret ${\bf C}_n$ as acting within the black hole interior, defined as the complement to the outside Fock space spanned by the $A_n$'s.

 The ${\bf C}_n$ are the AdS/CFT analogues of the Kraus operators
employed in \cite{vv,Preskill}.
The second relation in (\ref{kdef}) is the standard unitarity condition.
Note that (\ref{evol})  is a {\it linear map} from the space of Boulware  vacuum states $|\Psi_0\rangle$  to the space of equilibrium states $|\Psi\rangle$. Therefore, {\it the  internal ${\bf C}_n$ operators enjoy the same level of  state independence as the external $A_n$ operators}.

The requirement that $\li \Psi \ra$ defines a thermal state gives useful statistical information about the operators ${\bf C}_{n}$
\bea
\label{therm}
\la \Psi_0| {\bf C}^\dag_m {\bf C}_n \li \Psi_0\ra \is w_n \delta_{mn}\, .
\eea
with $w_n$ the Boltzman weight
\bea
\label{boltz}
w_n  = \frac{e^{-\beta E_n} }{Z}, \quad & & \quad \sum_n w_n = 1\, .
\eea
The relation (\ref{therm}) expresses the thermal equilibruim, or  detailed balance, between the external free field radiation modes and the internal state of the black hole.

\subsection{Interior Operators} 
\vspace{-3mm}

To construct the interior operators for equilibrium mixed states $\rho$, we need to introduce a mild form of state dependence.
Consider the same quantum quench scenario,  but now we prepare the CFT in some initial mixed state $\rho_0$ satisfying the Boulware vacuum conditions
\bea
\calO_\omega \rho_0 \is \rho_0\, \calO^\dag_\omega = 0,
\eea
and with Schmidt decomposition 
 \bea
 \rho_0 \is \sum_{\bar{i}} \, p_i \, \li\, \bar{i}\, \ra \la\, \bar{i}\, \ri.
 \eea 
 To this density matrix $\rho_0$, we associate a subspace of the CFT Hilbert space, defined via
\bea
{\cal H}_{\rm code} = \raisebox{-4pt}{$\mbox{\footnotesize $\bigoplus$} \atop \bar{i}$}\,  {\cal H}_{\bar{i}}
\eea
where ${\cal H}_{\bar{i}}$ denotes the $\calO_\omega$ Fock space 
built on top of the basis state $|\spc \bar{i}\spc \rangle$ in the Schmidt decomposition of $\rho_0$. 
We call ${\cal H}_{\rm code}$ the code subspace. 
In the following, we consider typical initial mixed states $\rho_0$, whose code subspace ${\cal H}_{\rm code}$ that is much larger than the $\calO_\omega$ Fock space, but much smaller than the total CFT Hilbert space.

Let  ${\bf P}$ denote the projection operator onto ${\cal H}_{\rm code}$. 
In other words ${\bf P}$ is the smallest projection operator that  leaves $\rho_0$ invariant and commutes with the $\calO_\omega$ oscillators
\bea
\label{pprop}
\spc{\bf P} \spc \rho_0 \spc =\spc \rho_0\,  {\bf P}\spc =\spc \rho_0, \qquad  \bigl[ \spc{\bf P}, \spc A_n \spc \bigr]\spc =\, 0.
\eea

After the quench, the CFT settles down in an equilibrium state
$\rho = e^{-i\tau H}\nspc \rho_0\spc e^{i\tau H} = \sum_{n,m} A_n {\bf C}_n \rho_0 {\mathbf C}^\dag_m A^\dag_m $.
Basic statistical reasoning shows that the matrix element of ${\bf C}^\dag_m {\bf C}_n$ between any pair of typical basis states in ${\cal H}_{\rm code}$ takes the form
\bea
\label{diag}
\la\, \bar{i}\, \ri {\bf C}^\dag{\!\!}_m {\bf C}_n\li \spc \bar  j\spc \ra   \is w_n\delta_{mn}\;  \delta_{\bar i \bar j}\, .
\eea
The diagonal form (\ref{diag}) arises via a simple mechanism: the operators $ {\bf C}^\dag_m$ and  ${\bf C}_n$ are large complex random matrices,
whose product decomposes as a sum of many terms with different phase factors. This sum
typically averages out to zero, except when there is constructive interference. Thermodynamics fixes the overall normalization.

The statistical property (\ref{diag}) plays a key role in the construction of the interior operators,
and holds with exponential accuracy,
provided that $\dim{H}_{\rm code} \ll \dim{\cal H}_{\rm CFT}$.
Though easily verified, it hides a small puzzle. The ${\bf C}_n$ all lower the
energy of the CFT state, and thus map a bigger to a smaller Hilbert space. Indeed, ${\bf C}_n$ is non-invertible on ${\cal H}_{\rm CFT}$. However, when restricted to any small subspace ${\cal H}_{\rm code}$, it becomes effectively invertible. 
This is the key observation that will allow us to build internal  $\calO^\dag_\omega$ oscillators, which, in spite of the fact that they lower the CFT energy, look like raising operators.

In \cite{vv},
motivated by the analogy with quantum error correcting codes \cite{Preskill}, we introduced recovery operators ${\bf R}_n$ and their hermitian conjugates ${\bf R}_n^\dag$ via
\bea
\label{recover}
{\bf R}_ n  = \, {\bf P} \, \frac{{\bf C}_n^\dag}{\sqrt{w_n\!\!}\;\,} \qquad ; \qquad
{\bf R}_n^\dag  = \, \frac{{\bf C}_n}{\sqrt{w_n\!\!}\;\,} \; {\bf P}\, .
\eea
 Using (\ref{diag}),
we deduce that  (up to exponentially small corrections) the recovery operator ${\bf R}_n$ acts on the equilibrium state $|\Psi\rangle$ via
\bea
\label{recprop}
{\bf R}_m \li \Psi \ra \is \sqrt{w_m\!\!}\; \,A_m \li \Psi_0 \ra
\eea
provided that $|\Psi_0\rangle \in {\cal H}_{\rm code}$. So by acting with the ${\bf R}_n$'s, one can recover the quantum information contained in the original vacuum state 
$|\Psi_0\rangle$.

Following \cite{vv}, we define the mirror $\widetilde{O}$ of the general operator $\hat{\cal O}$ specified in (\ref{matact}) as follows
\bea
\label{intone}
\widetilde{\cal O}\is    \sum_{m,n} \spc {\cal O}^*_{mn}\,  {\bf R}^\dag{\!\!}_n\spc {\bf R}_m 
\label{inttwo}
\eea
Note that these operators 
act linearly  on all of ${\cal H}_{\rm code}$.

\subsection{Mirror Property} 
\vspace{-3mm}

The mirror operator $\widetilde{O}$ represents the Hawking partner of the operator ${\cal O}$, the degree of freedom behind the black 
hole horizon that is entangled with ${\cal O}$.
In the notation (\ref{matact}), we can write the mirror property (\ref{mirrorone})-(\ref{mirrortwo}) as 
\bea
\widetilde{\cal O}\, A_\alpha \li \Psi \ra 
\is  A_\alpha\;
e^{-\frac \beta 2 H }\hat{\cal O}^\dag e^{\frac \beta 2 H}   \li \Psi\ra 
\label{mirrornew}
\eea
We would like to show that the operators (\ref{intone}) satisfy this mirror property for every state $\li \Psi\ra \in {\cal H}_{\rm code}$. 

The calculation is straightforward. Using the definitions (\ref{recover}) of the recovery property (\ref{recprop}), we deduce that
\bea
\label{help}
{\bf R}_n^\dag {\bf R}_m\li \Psi\ra  =  e^{\frac\beta 2 ( E_n-E_m) } A_m {\bf C}_n\li \Psi_0 \ra
\eea
With the help of this intermediate result, we compute
\bea
\widetilde{\cal O} \, A_\alpha \li \Psi \ra \is \sum_{m,n}{\cal O}^*_{nm} A_\alpha {\bf R}^\dag {\!\!}_n {\bf R}_m \,  
\li \Psi \ra\nonumber
\\[3mm]
    \is \sum_{m,n}  \spc e^{\frac\beta 2 ( E_n-E_m) } 
{\cal O}^*_{mn} A_\alpha A_m {\bf C}_n \,  
\li \Psi_0 \ra \nonumber \\[3mm]
\is \sum_{n}  A_\alpha\,\bigl( e^{-\frac \beta 2 H }\hat{\cal O}^\dag e^{\frac \beta 2 H}  A_n\bigr)\, {\bf C}_n\spc 
\li \Psi_0\ra\nonumber\\[3mm]
\is  A_\alpha\, e^{-\frac \beta 2 H }\hat{\cal O}^\dag e^{\frac \beta 2 H }  
\li \Psi\ra \nonumber
\eea
In the first line we used the definition (\ref{intone}) and that the ${\bf R}_n$'s commute  with $A_\alpha$. Next we use (\ref{help}).
The third step is a direct application of the notation~introduced~in~(\ref{matact}).  To arrive at the last line, we use equation (\ref{kdeco}).

The mirror property (\ref{mirrornew}) guarantees that the operators $\widetilde{\cal O}$ satisfy the same operator product algebra as the external efective QFT operators ${\cal O}$. For a more direct verification, and for a quantification of the corrections to this result, we refer to \cite{vv}. Note further that, unlike the PR construcion \cite{PR}, eqn (\ref{mirrornew}) is a derived property, rather than a postulate or definition. Moreover,
instead of just for a single state, (\ref{mirrornew}) holds for a large collection of states, namely all states $|\Psi\rangle = e^{-i \tau H}|\Psi_0\rangle$ with $|\Psi_0\rangle \in {\cal H}_{\rm code}$. This is why our definition of the mirror operators extends to mixed states of the form (\ref{mixed}).

\subsection{A Firewall Test} 
\vspace{-3mm}

A typical argument in favor of a firewall scenario  goes as follows. Suppose one has shown that a given equilibrium state $|\Psi \rangle$ has a smooth horizon. Now consider a new state obtained by acting with a unitary rotation
\bea
\li \Psi_{\rm new} \ra =  {\bf U}\spc \li \Psi \ra
\eea
where ${\bf U}$ commutes (to very high accuracy) with the external operators $\calO_\omega$ and with the Hamiltonian $H$. The new state then decomposes as
\bea
\li \Psi_{\rm new} \ra =  \sum_n A_n {\bf C}_{n, \rm new} \li \Psi_{0,\rm new}\ra
\eea
with $| \Psi_{0,\rm new}\rangle = {\bf U} |\Psi_0\rangle$ and ${\bf C}_{n, \rm new} = {\bf U}\, {\bf C}_n {\bf U}^\dag\, .$
It is then argued that, since ${\bf C}_{n, \rm new}$ looks different from the original interior operators
${\bf C}_n$, the new state can not also be regular at the horizon. 

Our construction (\ref{intone})
of the interior operators points to a clear loophole in this argument. Consider the  decomposition (\ref{kdeco}) of $|\Psi\rangle$. Since ${\bf U}$ is assumed to commute with the Hamiltonian and the $A_n$ operators, it must also commute with the operators ${\bf C}_n$. In particular,
${\bf U}$ commutes with the internal operators
\bea
{\bf C}_{n, \rm new} =  {\bf U}\, {\bf C}_n {\bf U}^\dag = {\bf C}_n
\eea
So in fact, by adopting the definition (\ref{intone}) of the mirror operators,  while including 
both $|\Psi\rangle$ and $|\Psi_{\rm new}\rangle$ in the code subspace, we see that both states are identified as states with a smooth horizon. 

Conversely, if one insists on creating a firewall state by acting on $|\Psi\ra$ with an operator ${\bf U}$ that does not commute with ${\bf C}_n$ operators, then ${\bf U}$ also does not commute with the Hamiltonian $H$. So  firewall states are non-equilibrium states \cite{PR,vv2}.

\subsection{Concluding Remarks} 
\vspace{-3mm}

We have shown how to construct CFT  operators that, within any given small subsector of the full CFT Hilbert space, have the correct semi-classical properties to be identified with free field theory modes in the black hole interior.   The formulas  (\ref{intone})-(\ref{inttwo}) define linear operators that satisfy the mirror property (\ref{mirrorone})-(\ref{mirrortwo}) for any state or density matrix within this subsector, and act as semi-classical free field oscillators up to exponential accuracy 
\be
\epsilon = e^{S_{code} - S_{\rm BH}}  \ll 1,
\ee 
where $S_{\rm code} = \log\dim {\cal H}_{\rm code}$. The mirror modes are the Hawking partners of the outgoing modes: they initially live on the left wedge of an eternal black hole. However, they can, without much trouble, be analytically continued to the region just behind the future horizon. 

The construction presented here follows the general prescription in \cite{vv}. As explained there, the success of the construction relies on the effectiveness of a `recovery operation', analogous to that used in quantum error correcting codes \cite{Preskill}. This recovery process works with high fidelity, provided the code subspace is small enough.

Though this result gives evidence against the firewall scenario, the main elements of the paradox remain in place.  In the CFT model for black hole evaporation, in which the asymptotic AdS modes can leak into an exterior system $X$, the CFT state will eventually  evolve into a 
highly mixed state that saturates the Bekenstein-Hawking bound.  At this point the recovery 
operation that underlies the semi-classical  reconstruction of the interior geometry  breaks down. 

This is not surprising.
The microscopic mechanism that brings out black hole information is unlikely to be explained by semi-classical physics. Information carrying radiation modes are not 
semi-classical: they correspond to errors in the recovery operation.
However, this is not a breakdown of the correspondence principle, but the result of a misguided  attempt  to capture the complete CFT Hilbert space in terms of a single semi-classical reality. Indeed, the lesson that seems to be emerging, is that the mechanism by which information can escape from black holes is related to the fact that quantum black holes can be in an incoherent superposition of many semi-classical states.  The construction of the interior operators must then involve a measurement, that selects one of the possible semi-classical realities.
 
 If one insists on writing the semi-classical interior modes as linear operators on the whole CFT Hilbert space, then this can be done as follows. Split the whole CFT Hilbert space into a direct sum of code subspaces
 \bea
 \label{decode}
 {\cal H}_{\rm CFT} =  \raisebox{-4pt}{$\mbox{\footnotesize $\bigoplus$} \atop k$} \; {\cal H}_{\rm code}^{(k)}
  \eea
Introduce an auxiliary external system (like the one mentioned a moment ago or by adding an observer) that becomes highly entangled with the CFT. In this set up, one can write linear operators of the type \cite{vv2}
\bea
\label{newop}
 \widetilde{\cal O} \, = \sum_k\; \widetilde{\cal O}^{(k)} \;  {\bf P}^{(k)}_{{}_{\!\! \rm X}}
\eea
where ${\bf P}_{{}_{\!\! \rm X}}^{(k)}$ projects on the particular subspace of ${\cal H}_{\rm X}$ that is correlated with the semi-classical CFT sector ${\cal H}_{\rm code}^{(k)}$. 
The operators (\ref{newop}) satisfy the  QFT operator product algebra and act as linear operators on the full CFT Hilbert space. In this sense, they are state independent. However, the operators (\ref{decode}) do depend on the entangled state of the combined system CFT plus X.
This type of state dependence can be physically acceptable.\footnote{This use of the exterior Hilbert space indeed looks reasonable given the physical fact that local Rindler modes at the horizon are virtual \cite{vv2}.  Local Rindler particles 
can only be made real by introducing an accelarated detector, and should therefore be thought of as modes created by the detector \cite{UW}.}

\medskip

\begin{center}{\bf Acknowledgement}
\end{center}
We thank Daniel Harlow, Juan Maldacena, Suvrat Raju and Edward Witten for helpful discussions.
The research of E.V. 
is supported by the Foundation of Fundamental Research of Matter (FOM), the European Research Council (ERC), and a Spinoza grant of the Dutch Science Organization (NWO). 
The work of H.V. is supported by NSF grant PHY-1314198.

\end{document}